\documentclass[conference]{IEEEtran}
\IEEEoverridecommandlockouts
\usepackage{cite}
\usepackage{amsmath,amssymb,amsfonts}
\usepackage{graphicx}
\usepackage{textcomp}
\usepackage{xcolor}
\usepackage{mathtools}
\usepackage{todonotes}
\usepackage{algorithm}
\usepackage{algpseudocode}

\newtheorem{definition}{Definition}[section]
\def\BibTeX{{\rm B\kern-.05em{\sc i\kern-.025em b}\kern-.08em
    T\kern-.1667em\lower.7ex\hbox{E}\kern-.125emX}}
\begin{document}

\title{Logic-based Task Representation and Reward Shaping in Multiagent Reinforcement Learning\\
{\footnotesize }
\thanks{This research was not funded by any organization.}
}

\author{\IEEEauthorblockN{Nishant Doshi}
\IEEEauthorblockA{\textit{Department of Robotics Engineering} \\
\textit{Worcester Polytechnic Institute}\\
Worcester, MA \\
ndoshi@wpi.edu}
}

\maketitle

\begin{abstract}
This paper presents an approach for accelerated learning of optimal plans for a given task represented using Linear Temporal Logic (LTL) in multi-agent systems. Given a set of options (temporally abstract actions) available to each agent, we convert the task specification into the corresponding Buchi Automaton and proceed with a model-free approach which collects transition samples and constructs a product Semi Markov Decision Process (SMDP) on-the-fly. Value-based Reinforcement Learning algorithms can then be used to synthesize a correct-by-design controller without learning the underlying transition model of the multi-agent system. The exponential sample complexity due to multiple agents is dealt with using a novel reward shaping approach. We test the proposed algorithm in a deterministic gridworld simulation for different tasks and find that the reward shaping results in significant reduction in convergence times. We also infer that using options becomes increasing more relevant as the state and action space increases in multi-agent systems.   
\end{abstract}


\section{Introduction}

Multi-agent Reinforcement Learning is the case of multiple agents trying to learn an optimal policy to a task in a stationary or non stationary environment. It is fundamentally harder than the single agent case because from the point of view of one agent, it's interacting with non stationary environment even though the actual environment is stationary as the other agents are concurrently moving and/or learning with it. When it comes to Multi-agent Reinforcement Learning, curse of dimensionality and thus exponential sample complexity makes it nearly infeasible for real world applications\cite{b1}. Also, one-step actions assumed in simplistic cases are never the case in the real world\cite{b2}. In case of multiple agents, each agent would have a set of capabilities/options which can be executed from a certain subset of state space.  These options (multi-timestep actions) have their own policies associated with them and may have stochastic terminating conditions i.e. there exists probability distributions of how long each option would last and what would be the terminating state. A sequential decision making problem with options can be formulated as a Semi-Markov Decision Process (SMDP)\cite{b3}. Additionally, it is always easier to define a collective task for the multiagent system rather than decomposing it into tasks for individual robots which is not only extremely challenging but also leads to a lot of assumptions in coordination, dynamic interplay of rewards and credit assignment among the robots. \cite{b4} 
When we consider centralized approach to multiagent planning, the problem boils down to an MDP or SMDP in case of options. In this work, we present a framework to logically represent a multiagent task as well as propose an approach to accelerate the learning in such systems.    

Linear Temporal Logic (LTL) is gaining popularity as a preferred framework for defining complex tasks. In an MDP, reward function is representative of the desired task to be completed. For a complex task, designing the reward function is a non-trivial problem in itself. The risk of using intuition or qualitative analysis to define a reward function is that it may not exactly represent the desired task. Proper estimation of reward function requires complex approaches like Inverse Reinforcement Learning. LTL provides a more natural way of defining a complex task. It also allows adding temporal dependency between different subtasks, for instance, a robot tasked with navigating from one room to another through a locked door, would have LTL subtask saying  "go to the door and unlock it only after finding and obtaining the key". In multi-agent systems, a new question arises-whether to define task centrally or decentralized tasks distributed to individual agents. In this context, we stick to centralized task definition and centralized control approach as the latter is more complicated and involves additional considerations such as task decomposition and coordination planning.\\
Given the options framework along with an LTL-based task specification, the following challenges lie ahead of us:
\begin{itemize}
    \item Choice of an appropriate reward function for the LTL-based task.
    \item Given a task specification in LTL, how to set/construct relevant options out of it so that the number of options is minimum to reduce the search space.
    \item If option is not known, it has to be additionally learned.
    \item In presence of multiple agents, further complications arise. Even though options for each agent itself is known, coupling between agents alter this model. Hence, unknown transition system arises as agents change their policies.
    \item Adapting an existing RL approach that would learn a correct-by-design policy to complete the task.
\end{itemize}
Keeping this in mind, options do reduce the search space by providing a set of primitive policies to complete the task. Especially in case of multi-agent learning/planning problems, this can prove to be a huge advantage and gain in computing speeds. 
\\

The organization of the paper is as follows: Section I
provides a general introduction and challenges one faces in multiagent reinforcement learning and the motivation for this work. Section II goes into defining any prerequisites required. Section
III formulates the problem and Section IV outlines the algorithm when everything is put together. Section V describes the experiments conducted to validate the algorithm. Section VI discusses the results for the experiments, while section VII describes the future scope of this work.

\section{PRELIMINARIES}

\begin{definition}
Options, in robotics, are higher-level, temporally extended behaviors (either handcrafted or learned) that simplify planning and execution. It’s a bridge between low-level control and high-level decision making. It is a tuple $O=\{\mathbb{I}, \pi, \beta\}$ where: 
\begin{itemize}
    \item $\mathbb{I}$ is a set of states from which it can be executed aka initiation set.
    \item $\pi$ is the policy executed.
    \item $\beta$ is the termination condition for the option. This can be a probability distribution or a rule when option ends.
\end{itemize}
An option is a generalized version of a single timestep action. It can generally last for multiple timesteps.
\end{definition}

\begin{definition}
A Labeled Semi-Markov Decision Process is a tuple $M=\{Q, \mathbb{O}, q_0, P, \mathbb{AP}, L\}$ where:
\begin{itemize}
    \item $Q$ is state set of the agent
    \item $\mathbb{O}$ is the option set available to the agent
    \item $q_0 \in Q$ is the initial state
    \item The transition function $P : Q \times \sigma \times Q \xrightarrow{} [0, 1]$ is defined such that $P(q, o, q') \in \{0, 1\}$ for states $q, q' \in Q$ and any option $o \in \mathbb{O}$.
    \item  $\mathbb{AP}$ is a finite set of atomic propositions and $L : Q \xrightarrow{} 2^{\mathbb{AP}}, \mathbb{AP}$ is a labeling function which assigns to each state $q \in Q$ a set of atomic propositions $L(q) \in \mathbb{AP}$ that are valid at the state $q$.
\end{itemize}
\end{definition}

\begin{definition}
A Product SMDP is a tuple 
\begin{center}
    $P=\{Q_1\times Q_2, \mathbb{O}_1\times\mathbb{O}_2, q_{10}\times q_{20}, P_X, \mathbb{AP}, L\}$
\end{center}
where:
\begin{itemize}
    \item $Q_X = Q_1\times Q_2$ is the augmented state set of two agents which is composed of individual states of each agent.
    \item $\mathbb{O}_X = \mathbb{O}_1\times\mathbb{O}_2$ is the option set available to the agent.
    \item $q_{10}\times q_{20} \in Q_X$ is the initial state.
    \item The transition probability $P_X$ function $P : Q_X \times \sigma \times Q_X \xrightarrow{} [0, 1]$ is defined such that $P(q_X, o_X, q'_X) \in \{0, 1\}$ for states $q_X, q'_X \in Q_X$ and any option $o_X \in \mathbb{O}_X$.
    \item $\mathbb{AP}$ is a finite set of atomic propositions and $L : Q_X \xrightarrow{} 2^{\mathbb{AP}}, \mathbb{AP}$ is a labeling function which assigns to each state $q_X \in Q_X$ a set of atomic propositions $L(q_X) \in \mathbb{AP}$ that are valid at the state $q$
\end{itemize}
\end{definition}

\begin{definition}
Stationary Memoryless policy is defined as a policy $\pi: Q\xrightarrow{}\mathbb{O}$ that outputs a distribution over option set based on the given state and doesn't depend on the history of states.
\end{definition}

\begin{definition}
Progress level function $P_l$ quantifies the degree of closeness to completing the specified task. 
\begin{center}
    $P_l: S\xrightarrow{}\mathbb{R}$
\end{center}
We use the approach mentioned in \cite{b5} to define the progress level function over the automaton states. The progress level acts as a heuristic for setting up the reward function as discussed later. 
\end{definition}

\begin{definition}
Let the sequence of states followed by a policy be $s_0, s_1, s_2,\dots$. The corresponding sequence of labels $q_0, q_1,\dots$ is called a word. Each task specification has a subset of words that satisfy it. Correspondingly, the Buchi Automaton of a specification has a subset of words that is accepted which is called the language of that automaton.  
\end{definition}

\section{Overview}

\subsection{LTL Formulation for Robotic Task Specification}
Linear Temporal Logic (LTL) provides a formal language to encode temporal and logical constraints in robotic task planning. It allows expressing complex behaviors such as sequencing, safety, and liveness requirements in a compact and verifiable manner. Let the workspace be abstracted into a finite set of atomic propositions $P={p_1, p_2, p_3...p_n}$, where each proposition corresponds to a specific location or task-relevant condition.
In the experiments sections, we consider a warehouse scenario with multiple delivery robots assigned. For a more simplistic case, a single robot tasked to pick up an object and drop it somewhere else can be specified in LTL as follows: (i) pick up an object at region $p_{pick}$, (ii) deliver it to $p_{drop}$, (iii) avoid entering restricted zones $p_{obs}$. An LTL specification is composed of the aforementioned atomic propositions and Boolean operators such as $\land$ (and), $\lnot$ (negation), and temporal operations on $p_x$. Some of the common temporal operators are defined as:
\begin{itemize}
    \item $G\phi:\: p_x$ is 'always' true i.e. true all future moments.
    \item $F\phi:\: p_x$ is 'eventually' true i.e. for some future moments.
    \item $X\phi: \:p_x$ is true the next moment.
    \item $p_1 \bigcup p_2:\: p_1$ is true until $p_2$ becomes true
\end{itemize}

This task can be expressed as the LTL formula:
\begin{center}
    $\phi = Fp_{pick} \land Fp_{drop} \land G\lnot p_{obs} \land G(p_{pick} \rightarrow p_{drop})$
\end{center}

This formulation encodes liveness (the robot must eventually visit pickup and drop-off points i.e. $Fp_{pick}$, $Fp_{drop}$), sequencing (the robot must only drop after picking up $G(p_{pick} \rightarrow p_{drop})$) and safety (restricted regions are never entered i.e. $G\lnot p_{obs}$). The resulting specification can be translated into a Büchi automaton and composed with the robot’s motion model to synthesize a provably correct controller. This approach enables robust, verifiable task execution even in dynamic or partially known environments.

\subsection{Centralized Multiagent Planning}
Centralized planning is a widely used paradigm for coordinating multiple autonomous agents through a single global decision-making process. In this approach, the joint configuration space of all agents is modeled explicitly, and a central planner computes coordinated trajectories or task assignments to achieve a common objective while ensuring inter-agent safety and optimality.
Let the set of agents be $\mathbb{A} = \{A_1, A_2, ... \}$, each with state $x_i \in X_i$ and options $o_i \in O_i$, the joint state of the entire system is $X = x_1 \times x_2 \times ... \times x_n$ and the central planner computes the following policy:
\begin{center}
    $\pi = X \implies o_1 \times o_2 \times ... \times o_n$
\end{center}
that drives all agents toward their goals while satisfying task and safety constraints such as collision avoidance.

Given this formulation, we end up having a transition system for each agent and an automaton for the task specification. The joint transition system for the resulting multi-agent system can be obtained by simply taking the product of each individual transition systems\cite{b6}. A tensor product can the be performed between the resulting joint transition system and the automaton. This essentially temporally augments each state in the joint transition system which results in a product MDP. A path search technique can then be used to find the optimal path in case of a reachability task or an optimal suffix cycle in some of the repeated satisfiability tasks. OPTIMAL{\_}RUN \cite{b7} shows some promising results for the latter and ROBUST{\_}OPTIMAL{\_}RUN algorithm improves upon it in terms of robustness. \\
Practical robotics applications may have huge MDPs associated with each agent and the above approach quickly becomes unfeasible. Some of the challenges underlying these applications include
\begin{itemize}
    \item The MDPs of each agent may be unknown or partially known.
    \item The agents may not be synchronized which would implicitly induce a larger search space while planning.
    \item If options framework is used, the agents actually get tightly coupled (both in state space and temporally) and the policy of one agent affects the termination states of options of other agents. It is very challenging to define such transition system so it is better to rely on sampling based methods. 
\end{itemize}
\subsection {Learning of MDP}
Keeping the above challenges in mind, we explore the possibility of learning the transition system by collecting transition samples.
This section runs through some methods for learning transition probabilities in MDP from the literature-
\begin{enumerate}
    \item Learning a Probably Approximately Correct (PAC) MDP. The central idea is to actively explore unknown state–action pairs while exploiting known dynamics when sufficient confidence is achieved. Early algorithms such as R-MAX and $E_3$ maintain optimistic models for unexplored states, encouraging exploration that drives learning. Once the agent has gathered enough samples to estimate transition dynamics and rewards accurately, it plans according to the learned MDP. Plan correctness depends on the approximation quality which implies large number of simulations required for a good approximation.\cite{b12}
    \item Learn transition probabilities using Maximum Likelihood Estimation (MLE): Collect $\{X, O, X'\}$ tuples and use frequencies to estimate the transition probabilities. As more and more samples update this estimate, the transition probabilities converge to true transition model.
\end{enumerate}
Using model-based methods would rather require a large number of samples per iteration which would grow exponentially with increasing number of agents and also requires adaptation to changing policy of other agents. Next we explore model-free methods for learning a plan to achieve an LTL based task.\\

\subsection{Model-free approach as a way to avoid learning the transition model}
A few model-free approaches do exist in the literature. 
\begin{enumerate}
    \item $\tau$-Q learning assumes finite horizon tasks defined in Signal Temporal Logic. It considers history of states as states of ($\tau$-MDP) and length of history depends on the task specification(number of samples required to resolve requirements of a task $\phi$). Reward function is elegantly mapped to task definition and is approximated to be use in Q learning. \cite{b8}
    \item Q learning can be applied directly, and states of MDP and automaton are coupled together on-the-fly similar to the approach presented in this paper. 
    
\end{enumerate}

The next section proposes a method that doesn't require learning of model and performs piecewise product operation of the unknown joint transition system and the task automaton on-the-fly. The reward function is set up so as to guide the agent to complete the task inspired by reward shaping. \\

\section{Logically Constrained guided Multiagent Q Learning}
We present a model free approach to learning a Q function for a task specified in LTL. It's important to note that it doesn't involve explicitly constructing a product MDP. Instead the states of the MDP are temporally augmented(combined with automaton state) on-the-fly as the samples come. Exploration also involves taking actions that would lead to a next state acceptable to the task automaton.\\
This leads to model free algorithm to learn plans to temporally defined task. for the initial part, the MDP is assumed to be deterministic. But stochastic case can be dealt with, with slight modification as discussed later. \\
The algorithm uses an indirect reward function defined based on the progress of the agent (Algorithm 2) in the completion of the task. This plays a big role not only in the speed of convergence but the convergence itself. We further add that the reward function should be a potential function of progress level in lines of \cite{b9} to retain policy invariance and optimality. Options and the logical task definition along with progress reward function together guide the agents into completing their task faster hence the name. \\

\begin{center}
\begin{minipage}{8.5cm}
\begin{algorithm}[H]
\caption{LCgMQL: Logically Constrained guided Multiagent Q-Learning}
\begin{algorithmic}[1]
\State \textbf{Input}: Environment env, Exploration probability $\epsilon$, discount factor $\gamma$, learning rate $\alpha$, size of replay memory N, number of replays K, Task converted to Buchi Automaton A with start state , set of all possible joint options $JO$
\State $Q\xleftarrow[]{}\emptyset\;\:\forall S_O$      //Initialize Q values
\State $M \xleftarrow[]{}\emptyset $    // Experience Buffer
\State $q\xleftarrow{}q_0$  //Assign current automaton state for book-keeping
\State $s\xleftarrow{}env.current\_state$
\State
\State /// Start Q Learning iters
\While{not reached stopping criterion}
        \State c = 0    // iteration counter
        \For {t = 1, 2, \dots }
        \State /// Joint Options executable in current state s
        \State $eO\xleftarrow{}JO\setminus Non-executable(JO,s)$
        \State /// Gathering Permissible joint options for current state
        \State $A_p =\begin{cases}
        (o_{active}\otimes o) \:\forall o_{active}\:for\:o\in eO\setminus{o_{active}} \\
        eO\:if\:Active\:Options=\emptyset
        \end{cases}$
        \State /// Choose option with exploration
        \State $a_t =  \begin{cases}
         argmax_a\:Q(s_t, a) \:with \:probability\: 1-\epsilon \:\& \\a\in A_p\\
                random \:option\:from\:A_p\: with\:probability\:\epsilon.
        \end{cases}$
        \State /// Rollout/simulate 1 step of chosen options
        \State simulate $\{s_t,a_t\}\xrightarrow{}s_{t+1}$  
        \State /// Check if transition leads to state label acceptable to A 
        \If {$Label(s_{t+1}) \:exists\: in\: A.transitions(q)$}
            \State $prev\_q\xleftarrow{}q$
            \State $q\xleftarrow{}  A.next\_state(q)$
            \State $E \xleftarrow{} \{(s_t,prev\_q), a_t, (r_t + shaped\_reward(s_{t+1},q,A)), (s_{t+1},q)\}$ // Collect experience; also refer to \textbf{Algorithm 2}
            \State $M \xleftarrow[]{} Push(M,E)$
        \EndIf
        \State $K\xleftarrow{}Sample\: N\: experiences\: from\: M$
        \State $updateQ(Q, K, \gamma, \alpha)$
        \EndFor
        \State  c = c + 1
\EndWhile
\State \textbf{Return} $\pi*$
\end{algorithmic}
\end{algorithm}
\end{minipage}
\end{center}

For Algorithm 2, the nodes of the automaton have to be annotated with progress level as follows-
\begin{enumerate}
    \item Add initial state to the openlist. Set progress level to 0
    \item Pop all states from openlist
    \item For each popped state, find all strongly connected components in BA that contain it.
    \item Annotate all strongly connected components with current progress level.
    \item Add all the states in the boundary** of strongly connected components to the openlist of states. 
    \item Increment progress level; go to 2 
\end{enumerate}

\begin{minipage}{8.5cm}
\begin{algorithm}[H]
\caption{shaped\_reward}
\begin{algorithmic}[1]
\State \textbf{Input}: MDP state $s_{t+1}$, automaton state $q$, automaton A
\State $l\xleftarrow{}BA\_Progress\_Level(q,A)$ \cite{b11}
\State return (something $\propto \: l$)
\end{algorithmic}
\end{algorithm}
\end{minipage}
\\\\
Essentially, the above algorithm ensures that progress is made only when the adjacent nodes are not strongly connected which also makes sense intuitively. If there is a way of coming back to the node from which we move forward, the progress is reversed or in other words there is no progress and hence all strongly connected components are annotated with the same progress level. **Boundary in this context means all the nodes directly adjacent to a strongly connected component which are not a part of that strongly connected component. This would make the boundary nodes a part of the next progress level which can be observed in the next iteration as the control jumps to 2. Once we have the annotations, they can be directly used in Algorithm 2.
\\
The following assumptions hold:
\begin{itemize}
    \item Option definition (Initiation condition, Termination condition, Internal policy) is known for all options available to all the agents.
    \item Termination condition is deterministic.
    \item The states are fully observable.
    \item The model of environment (MDP for each agent) is not known. 
    \item The environment states are statically labeled for each agent.
\end{itemize}

Line 16 is similar to $\epsilon$-greedy action selection in Q-Learning. Once the joint actions/options are chosen, the next step is to evaluate them. A single step simulation is done (Line 18) to verify that the label produced by the next state is acceptable to the task automaton. This is done in Line 20. All valid transitions are stored as experiences with the MDP states augmented with automaton states. The reward function comes from \textbf{Algorithm 2} which returns a reward proportional to the progress level of the new state in automaton. Line 14 is crucial in capturing the effects of options. It is important to note that each sample is just 1 step and each option not hitting its terminating condition is set as active. While choosing joint options in the next iteration, line 14 ensures that no option selection occurs for agents with active options. Thus active options remain as is in the new joint option chosen and for all other agents, the option selection occurs as usual from a set of available executable options. For all experiences having active options, once the option/s terminate, Q values of all respective state-action pairs are updated such that the reward received at the end of a joint option execution is cascaded along the trajectory for which the option/s were active. Additionally, replay memory enables updating in batches and also enables quick adapting in lowly dynamic environments but no experiments were performed in dynamic environment. \\
As long as the final automaton state is reachable from the augmented initial state of the agents, there exists at least one solution path. Additionally, the convergence proof follows on the lines of traditional Q learning and the following conditions must hold:
\begin{itemize}
    \item Each state-option pair is visited infinite number of times.
    \item Step size sequence $\alpha_{t_1},\alpha_{t_2},\dots$ follows: $\sum_t \alpha_t =\infty$ \& $\sum_t (\alpha_t)^2 <\infty$ 
\end{itemize}
\section{Experimental Results and Discussion}
During experiments, we found that convergence speed depended on the exploration parameter $\epsilon$ and maximum trajectory length before everything was reset. Also, discount factor would play a huge role, as the reward is fundamentally a multistage sparse reward and not continuous. This effect can be explained with the following figures below. 
\begin{center}
\includegraphics[width=9cm, height=5.5cm]{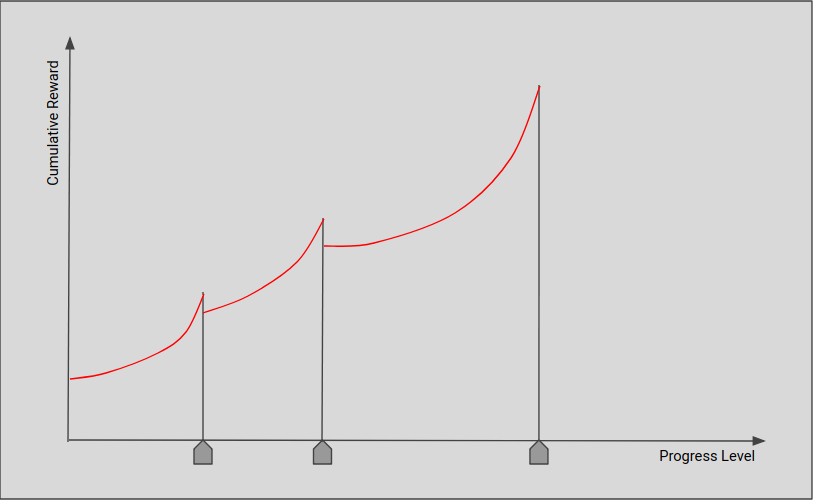}\\
\textbf{Figure 2.5a: Discount factor is very low; reward signals decay faster backwards}
\includegraphics[width=9cm, height=5.5cm]{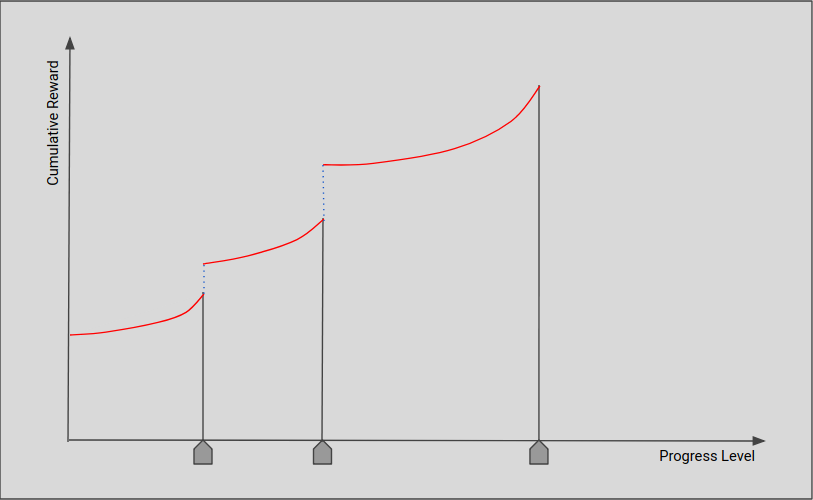}\\
\textbf{Figure 2.5b: Discount factor is very appropriate; reward signals decay as desired}
\end{center}
In Figure 2.5a , one issue that may arise is the agent, instead of progressing in automaton, would learn to stay at the same progress level because it sees less cumulative reward for reaching the next progress level. 
In Figure 2.5b, choosing appropriate $\gamma$ value actually solves this issue and the agent would learn to progress correctly. 
One way around this is to not use discounting at all as long as we know that we are working with finite horizon problems. The next section runs through the experimental results on running this algorithm.

\subsection{Experiments: Gridworld-Reach-Avoid Task}
This section discusses the results on running \textbf{Algorithm 1} for the aforementioned Gridworld-Reach-Avoid task. All the notations and definitions follow as is. \\
We used Spot platform and its Python bindings for translating the LTL task definition to a Buchi Automaton. A custom wrapper was used for additional manipulability of resulting graphs\cite{b12}.   \\
The reward function was rather simple: progress level times 50. This multiplier was chosen after fixing the discount factor to 0.9 and examining how the utilities decay per step. \\
To start with, the algorithm was first tested for a simpler case where each agent has five primitive actions- Up, Down, Left, Right and Stay. Given a task specification the agents need to figure out a plan to reach their respective goals safely. The following plan was obtained for the test configuration.   \\
\begin{center}
\includegraphics[width=8cm, height=8cm]{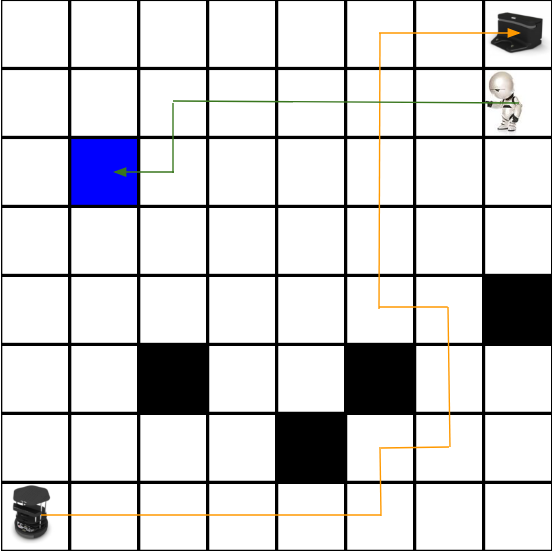}\\
\textbf{Figure 2.6: Plan generated by the algorithm for 4 primitive actions}
\end{center}
It can be observed from the figure that the path for the turtlebot is not optimal. Though we believe that running the algorithm a bit longer would eventually result in the most optimal paths for both the agents. The training lasted for 12000 steps with 32 updates per step(batch of 32 sampled from replay memory) and took around 203 seconds.\\
Next we consider the case of options. In this case we use four primitive actions as options as well as the go-to-goal and obstacle avoid options as described in the previous section. The following plan was generated by the Algorithm 1:
\begin{center}
\includegraphics[width=8cm, height=8cm]{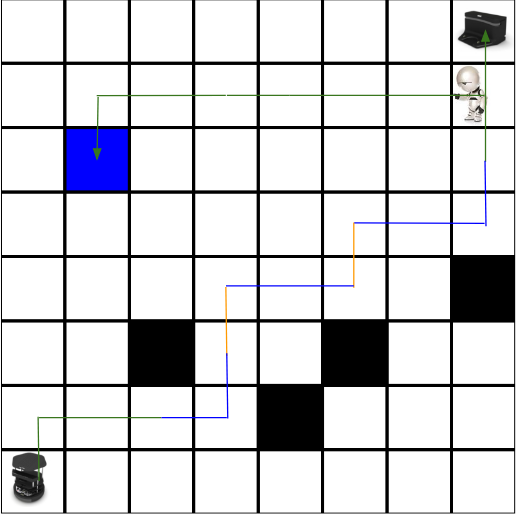}\\
\textbf{Figure 2.7: Plan generated by the algorithm for 4 primitive actions + go-to-goal and collision-avoid options}
\end{center}
In the above figure the following color coding is used:
\begin{itemize}
    \item Blue arrows are for the four primitive actions- Up, Down, Left, Right
    \item Green arrows are for go-to-goal option
    \item Yellow arrows is for obstacle-avoid option
\end{itemize}
The solution found was better than the previous case with no options. But it is important to note that none of the task constraints have been violated by the plan and also the utilities are appropriately propagated backwards if an option remains active. The training lasted for 12000 steps with 32 updates per step(batch of 32 sampled from replay memory) which took around 240 seconds.
\\
\begin{center}
\includegraphics[width=9cm, height=8cm]{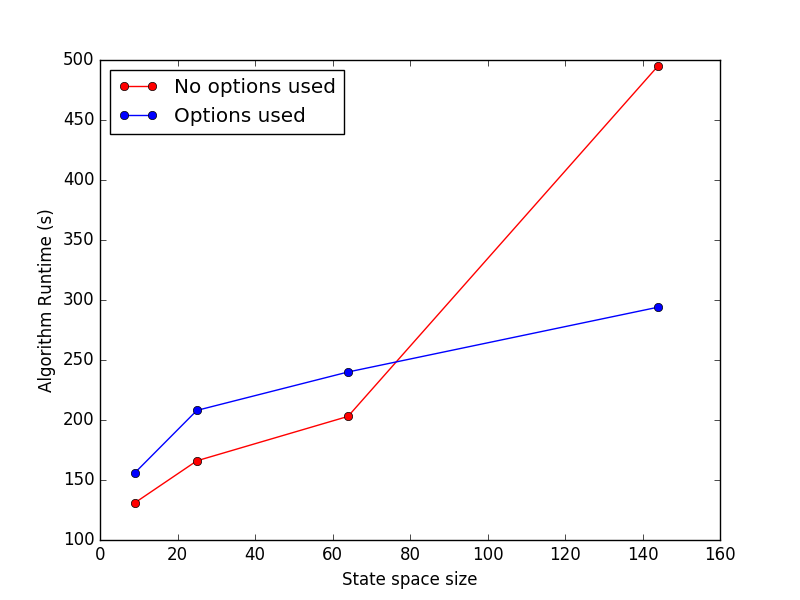}\\
\textbf{Figure 2.8: Plot of state space size vs algorithm runtimes}
\end{center}
It can be seen in Fig. 2.8 that it is more practical to use LCgMQL with options as the state space size increases. For higher grid sizes it took a lot more training iterations for the algorithm to converge. The utility propagation step in case of options leads to huge speed ups during learning process.

\begin{table}[h!]
\centering
 \begin{tabular}{|c c c c c|} 
 \hline
  & 3x3 & 5x5 & 8x8 & 12x12\\ [0.5ex] 
 \hline\hline
 No Options & 131 & 166 & 203 & 495 \\ 
 \hline
 With Options & 156 & 208 & 240 & 294 \\
 \hline
\end{tabular}
\caption{Algorithm Run-times (s) for different Grid sizes.}
\label{table:1}
\end{table}
Table 2.1 clearly shows the exponential blowup in algorithm run-times as the grid size increases and no options are used. It is important to note that for 12x12 grid case, convergence was achieved with a lot more iterations (30000) in contrast to the case with enabled options which took only 15000 iterations. In spite of increased number of iterations, optimality was not consistent. As the grid size increases it becomes more relevant to use options

\section{Conclusion and Future Work}
This work shows the possibility of combining high level logical specifications and model-free reinforcement learning. It presents a more feasible centralized multi-agent reinforcement learning approach which can be used given a complex task specification in the form of Linear Temporal Logic. It has always been a challenge to come up with a reward function the captures the task desired especially in case of multiple agents and this approach greatly reduces the difficulty of ‘designing’ reward functions. A direct next step would be to explore the direction of probabilistic tasks defined as Computational Tree Logic. This would open up avenues to mission based planning in  partially observable systems. Another direction is to extend this work for continuous state space to assess its viability. There exist Hybrid Control approaches that employ feedforward and feedback schemes at higher and lower levels respectively provided the models are known\cite{b13}. But in case of partial models, this work may help set a direction to solving the single/multiagent planning problem.

\end{document}